\documentclass[%
 reprint,
 amsmath,amssymb,
pra,
]{revtex4-2}

\usepackage{graphicx}
\usepackage{dcolumn}
\usepackage{bm}

\usepackage{systeme}
\usepackage{verbatim} 
\usepackage{wasysym}

\begin{document}


\title{Transformation of the Talbot effect  in response to  phase disorder}

\author{Ilia Mosaki}
\affiliation{%
Department of Physics,  Lomonosov Moscow State University,
 Moscow, 119991, Russia
}%

\author{A.  V.  Turlapov}
 \email{a\_turlapov@mail.ru}
\affiliation{
A.V. Gaponov-Grekhov Institute of Applied Physics of the Russian Academy of Sciences, \\
  Nizhniy Novgorod, 603950,  Russia
}%
\affiliation{
Federal State Unitary Enterprise \textquotedblleft  Russian Metrological Institute of Technical Physics and Radioengineering\textquotedblright,  \\ 
Mendeleevo, 141570, Moscow region, Russia
}%

\date{\today}

\begin{abstract}

Bose--Einstein condensates initially arranged  in  a long chain freely expand and interfere.
If the initial phases of the condensates are identical,    the initial density distribution   is  restored periodically  during the  expansion,   giving rise to the Talbot effect.
Even a slight disorder in the initial phases 
 leads to a transformation of  the interference pattern.
In response to the  phase disorder, the spectrum of the spatial density distribution acquires  peaks that are absent  in the case of identical phases.
    We derive an analytical expression for  the  spectrum of the spatial density distribution   for  an arbitrary phase disorder.
We show that the new peaks emerging due to the phase disorder originate from  pairwise interferences of the condensates.
The positions of these peaks coincide with the wave vectors of the density modulations (wavelets) generated by such pairwise interferences.
The absence of these peaks,   when the initial phases are identical,  is explained by  the mutual destruction of the overlapping wavelets during their summation.

\end{abstract}

\maketitle

\section{\label{sec:level1} Introduction}

In  optics,  the Talbot effect results from the Fresnel diffraction of light by a periodic grating.
At distances $n Z_d  / 2$ from the grating,    the diffraction produces  copies of the initial intensity distribution \cite{Talbot,  Rayleigh_Talbot_effect}.
 Here  $n\in \mathbb{N}$ and $Z_d = 2d^2 / \lambda $ with $d$ being the grating period  and  $\lambda$ being the light wavelength.
 If $n$ is even,   the copies exactly match the initial  intensity distribution.
 If $n$ is odd,   the copies  are laterally shifted by  $d/2$  relative to the initial grating.

Apart from the standard Talbot copies,  the Fresnel diffraction can also produce more complex spatial intensity structures.
For example,  at  distances  $ \frac{1}{2}Z_d  (n + \frac{p}{ q})$,   where $p < q$ are coprime positive integers,  
the spatial period of the intensity distribution can be reduced by a factor of $q$ compared to the initial period \cite{Hiedemann1959, Winthrop1965}.  This phenomenon is known as the fractional Talbot effect.
At  distances  $\frac{1}{2} Z_d ( n + s)$,  where  $s$ is an irrational number,   the fractal Talbot effect  may arise \cite{Berry1996}.
In this case, the spatial intensity distribution is a function of the position that is continuous everywhere but differentiable nowhere.
Together, the ordinary, fractional, and fractal Talbot effects form an intricate spatial intensity structure known as the Talbot carpet \cite{Berry2001}.
The Talbot carpet is connected to the number theory \cite{Berry1996}  and can be applied to a factorization of integer numbers \cite{TalbotEffectFactoring1996}.

Analogues of the optical Talbot effect have been  identified in acoustics \cite{AcousticTalbot1985,USoundTalbot2017eng},  plasmonics \cite{PlasmonTalbot2009},  spintronics \cite{SpinwaveTalbot2012},  polaritonics \cite{Gao2016},  and atomic physics \cite{PritchardTalbot1995}.
An analogue  of the fractional Talbot effect has  been  demonstrated in atomic physics \cite{Nowak1997}.  
In addition to  the spatial Talbot effect,  the temporal Talbot effect has been observed,  for instance, during the propagation of a light pulse train through a fiber  \cite{Jannson1981,  Andrekson1993}.
In the physics of ultracold atoms and molecules,  the temporal Talbot effect has been observed during the free expansion  of Bose--Einstein condensates released  from a one-dimensional optical lattice \cite{ FluctPhaseTalbot2017, RandomPhaseInteference2019}.  
If interparticle interactions during the expansion of the condensates are absent,  the replication of the initial density distribution  occurs at expansion times   $ n T_d/2$,  where $n \in \mathbb{N}$,  $T_d = \frac{md^2}{\pi\hbar}$,  with $m$ being the mass of a single particle within the condensates.

Defects and fluctuations in the initial condition of the Talbot effect can arise for various reasons.
For example,  in a diffraction grating,    defects may result from careless handling.
In an array of solid-state lasers,   fluctuations of the laser intensities  can originate from pumping inhomogeneities  \cite{Kono2000}. 
When a single defect is present  in the amplitude or  phase of the initial condition,   self-healing occurs at the Talbot  distances  \cite{Dammann1971, Kalestynski1978, Lu2005, Wang2010, Teng2015}: the defect diminishes in the replicas of the initial condition,  with  suppression becoming more pronounced  at greater distances.
This  self-healing property has been proposed as a mechanism for generating defect-free structures in nanolithography \cite{Urbanski_2012, Li_2013}.

Beyond the isolated defects,   the self-healing has also been demonstrated  for several specific types of  fluctuations  distributed across  the entire initial condition of the Talbot effect.
 For instance,  in the diffraction grating,  the self-healing has been observed when all  slits  exhibit fluctuations in the shapes or positions \cite{Smolinska1978}.
Similarly,  when the Talbot effect initial condition is formed by an array of distinct wave sources,  the self-healing occurs  when the amplitudes across all sources fluctuate \cite{Takai_amplitude_fluctuations_1995}.
As a result of this self-healing behavior, the spatial intensity distribution at the Talbot distances preserves the same periodicity as in the ideal, defect-free case. 
Consequently, the spectrum of the spatial intensity distribution consists of discrete peaks at wave numbers $k = 2\pi l /d$ $\left( l \in\mathbb{Z} \right)$.

In contrast to the  fluctuations of the amplitudes,   fluctuations of the phases across all wave sources  qualitatively change the Talbot effect.
This phenomenon  was experimentally observed in a chain of freely expanding Bose--Einstein condensates \cite{RandomPhaseInteference2019}.
In such chains,   the phases of the condensates can fluctuate due to  interactions of the condensates with thermal particles or due to the quantum uncertainty \cite{BoseChainFluctPitaevskii2001}.
Once the phases of the condensates become disordered, the spectrum of the spatial density distribution acquires additional peaks that are absent when the initial phases are identical.
If  interparticle interactions during the  expansion of the condensates are absent,  these peaks occur at wave numbers $k = \pi l T_d  / (t d)$,  where $t$ is the expansion time and $ l \in  \mathbb{Z} / \{   0 \} $. 
When the initial phases are only partially disordered, both the disorder-induced peaks and those associated with  uniform phases are simultaneously present in the spectrum.
Under  completely disordered phases,  only the disorder-induced peaks remain.

In comparison with the non-interacting case,  interparticle interactions during the expansion introduce quantitative corrections to the interference pattern of  Bose--Einstein condensates.
Specifically,  
when the phase fluctuations are absent, 
 the interactions shift the Talbot revival period $T_d$ and  
 induce a distortion of the spatial density distribution \cite{Hollmer2019,  Fansu2024}.
In the presence of the phase fluctuations, the interactions cause slight shifts in the positions of the disorder-induced peaks,  and deform  the peak shapes \cite{RandomPhaseInteference2019,  GPEinterference2024_eng}.
Since  the qualitative structure of the spatial density spectrum remains unchanged in the presence of the interactions,  it can be correctly captured in the non-interacting limit.
We therefore consider Bose--Einstein condensates without the interactions in this paper.

 In this paper,  
  for a chain of Bose--Einstein condensates expanding without interparticle interactions,  
  we derive  an analytical expression for  the  spectrum of the spatial density distribution   under  an arbitrary phase disorder.
We show that the spectrum peaks induced by the phase disorder originate from  pairwise interferences of the condensates in the chain.
The wave vectors of these peaks coincide with the wave vectors of the density wavelets generated by such pairwise interferences.
We explain the absence of such peaks in the case of identical initial phases as resulting from the mutual destruction of  the density wavelets  during their summation.
Furthermore,  we demonstrate that  the peaks originating  from the pairwise  interferences appear in the spatial density spectrum for any condensate lattice, regardless of its geometry.

In addition to the Fresnel diffraction regime  where the Talbot effect occurs, the analytical expression for the spatial density spectrum derived here is also valid in the Fraunhofer diffraction regime.
According to this expression,  in the Fraunhofer regime,  the phase disorder  leads only to quantitative changes in the spectrum, in contrast to the qualitative changes in the Fresnel regime.
Specifically, phase fluctuations modify the spectrum by altering the heights of the peaks.
In the long-chain limit,  our spectrum expression further predicts that the relative peak heights are completely determined by the phase correlation function of the condensate chain. 
This result is consistent with the experimental and theoretical analysis of the spatial density spectrum in the Fraunhofer  diffraction regime reported in Ref. \cite{Wang2012}.

Section \ref{sec:model} describes the interference model for a chain of Bose--Einstein condensates and provides examples of density calculations.
Section  \ref{sec:spectrum}  derives the formula for the spatial density spectrum under an arbitrary phase disorder.
Section \ref{sec:explanation} presents the qualitative  explanation of the  spectrum peaks emerging due to the phase disorder.
In Section \ref{sec:general_lattices},   we consider  interference of a condensate lattice with arbitrary geometry. 
In Section \ref{sec:Fraunhofer},   we compare  the interference of condensates in the Fresnel regime with the  interference in the Fraunhofer regime.
Finally, Section \ref{sec:conclusion} provides the conclusion.

\section{Model of interference of a chain of Bose condensates \label{sec:model} }

Following experiments on ultracold atoms  and molecules \cite{RandomPhaseInteference2019, Anderson1998,  Orzel2001,  Burger2001},  a chain of Bose--Einstein condensates is prepared in the periodic potential of a one-dimensional optical lattice.
Near each minimum,  the potential is  approximately harmonic.   Therefore,   the wave  function of the condensate chain can be represented as
\begin{equation}
  \Psi( \mathbf{r} ,  t=0)   = \frac{N_0^{1/2}}{(2\pi)^{1/4}\sigma^{1/2}}  \sum\limits_{j=1}^M e^{-\frac{{(z-jd)}^2}{4\sigma^2}} e^{i\varphi_j}\chi(\mathbf{r}_{\perp}, t=0)  ,  
  \label{initial}
\end{equation}
  where $\sigma $ denotes the width of each condensate along the chain axis,   $d \gg \sigma $  is the chain period,    and $\varphi_j$   is the phase of the $j$-th condensate.
The chain consists of  $M$ condensates,  each containing   $N_0$ particles.  
  The function $\chi(\mathbf{r}_{\perp},  t=0)$ describes the dependence of the  wave function    $\Psi(\mathbf{r}, t=0)$ on the coordinates perpendicular to the chain axis.

In the absence of the interparticle interactions,    the evolution of the wave function $\Psi(\mathbf{r},t)$ during a free expansion of the condensates  from the optical lattice is governed by the free-particle Schrödinger equation.
The interference of the condensates during the expansion occurs along the chain axis,  therefore the interference is  completely determined by the axial part $\psi(z,t)$ of the wave function $ \Psi(\mathbf{r},t) = \psi(z,t)\chi(\mathbf{r}_\perp,t)$. 
For the axial part  of the wave function,  the solution of the Schrödinger equation is
\begin{equation}
  \psi ( z ,t)  = \frac{N_0^{1/2}}{(2\pi)^{1/4}\sigma^{1/2}\sqrt{ 1 + i\omega t }}\sum\limits_{j=1}^M e^{-\frac{1}{4\sigma^2}\frac{{(z-jd)}^2}{1 + i \omega t}} e^{i   \varphi_j  } ,
\label{wave_function}
\end{equation}
where $m$ is the mass of a single particle of the condensates,
$\omega =\hbar  / (2m\sigma^2) $.

\begin{figure*}
\includegraphics[scale=0.97]{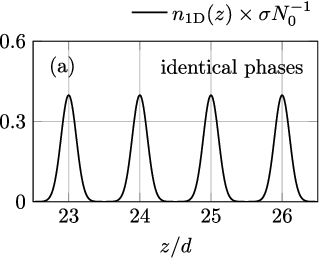}
\hspace{0.6cm}
\includegraphics[scale=0.97]{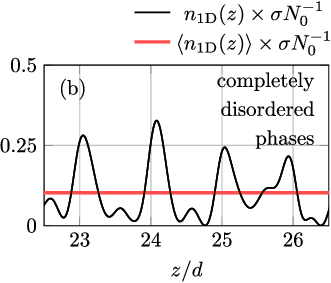}
\caption{
One-dimensional density profile of a freely expanding chain of Bose--Einstein condensates at time $t=T_d$.
The  spatial region where the Talbot effect can occur is shown.
The initial phases of the condensates are (a) identical,  (b) completely disordered.
}
\label{fig:density}
\end{figure*}

According to expression \eqref{wave_function},  
at  times $  t \ll   M d / (\sigma\omega) $,   the width of a single expanding condensate $ \sigma \sqrt{1 +  \omega^2 t^2}$ is much smaller than the total   length  of the chain  $Md$, 
and   there exists a spatial region where the one-dimensional density  $n_{1\text{D}}(z,t) = |\psi(z,t)|^2$ is the same as that produced by an infinitely long chain of condensates  \cite{Gammal2004}.
The replication of the initial density  occurs within this  region at  times $ nT_d /2$ $\left(n \in \mathbb{N}  \right)$.
An example  of such replica   at time $t=T_d$  is shown  in Fig.  \ref{fig:density}(a).

A disorder in the  initial phases  alter the shape of  the density modulations  produced by the interference of the condensates.
Within the spatial region where the Talbot effect occurs,  these modulations lose their exact periodicity although their contrast remains high.
An example of the one-dimensional  density $n_{1\text{D}}(z,t)$ under disordered initial phases is shown in Fig. \ref{fig:density}(b).
The density was calculated using equation \eqref{wave_function} with the initial phases  treated as uncorrelated random variables: $ \langle e^{i(\varphi_j - \varphi_{j+1})} \rangle = 0$, where $\langle ... \rangle$  denotes averaging over  calculation repetitions.  
Due to the randomness of the initial phases, the specific shape of the density profile varies between the repetitions of the calculation.
Averaging  $n_{1\text{D}}(z,t=T_d)$ over these repetitions causes the density modulations to vanish,  as illustrated by the red line in Fig. \ref{fig:density}(b).

\section{Spectrum of the spatial density distribution}

\label{sec:spectrum}

\begin{figure*}
\center
\includegraphics[scale=0.85
]{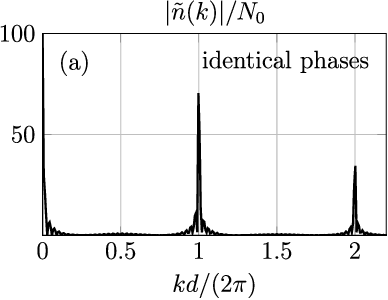}
\hspace{0.8cm}
\includegraphics[scale=0.85]{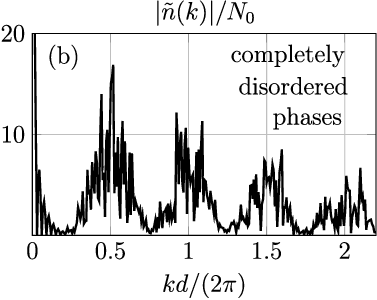}
\caption{\small
Amplitude   of spatial density spectrum $\left| \tilde{n}(k,t) \right|$ for  a chain of   $M=100$   Bose--Einstein condensates   at time  $t = T_d$.
The initial phases of the condensates are (a) identical,  (b) completely disordered.
}
\label{fig:spectrum1}
\end{figure*}

The  density modulations resulting from the interference of the condensates with the completely disordered initial phases,  shown in Fig. \ref{fig:density}(b), exhibit a remarkable property.
Although the specific shape of the modulations varies from one experimental realization to another, the amplitude of the spatial spectrum $ \left| \tilde{n}(k,t) \right|$, 
where \begin{equation}
\tilde{n}(k,t) = \int \text{d}z n_{1\text{D}}(z,t) e^{ - ikz}, 
\label{spectrum_definition}
\end{equation}
remains reproducible \cite{RandomPhaseInteference2019}.
Moreover,   the spectrum amplitude $ \left| \tilde{n}(k,t) \right|$  under disordered phases differs qualitatively from that obtained with uniform phases.
Figs.  \ref{fig:spectrum1}(a), (b) show numerically computed examples of $ \left| \tilde{n}(k,t) \right|$ at $t=T_d$   for uniform and completely disordered initial phases, respectively.
The figures demonstrate that, for completely disordered phases, the spectrum amplitude consists of several well-defined peaks that are absent when the phases are identical.

We now derive  an analytical expression for the  spectrum amplitude  under an arbitrary  disorder in the initial condensate phases.
At first,  we calculate  integral \eqref{spectrum_definition} with $n_{\text{1D}}(z,t)= | \psi(z,t)|^2$ obtained from  wave function \eqref{wave_function}:
\begin{multline}
\tilde{n}(k,t)
 = N_{0} e^{-k^2\sigma^2/2}  \\
 \times \sum\limits_{n = 1-M}^{M-1} \exp\left(  -\frac{d^2}{8\sigma^2}   \left( \frac{ktd}{\pi T_d}  - n \right)^2    \right) f_n(k) , 
  \label{spectrum}
\end{multline}
where
\begin{equation}
 f_n(k) =  
e^{-ikd|n|/2}   \sum\limits_{l=1}^{M-|n|} e^{-i \text{sgn}(n)( \varphi_l - \varphi_{l+|n|}  )}  e^{-ikld }  ,
\label{f0}
\end{equation}
with $\text{sgn}(n) = +1$ if $n \geqslant 0$,  and $\text{sgn}(n) = -1$ if $n < 0$.

Before taking the absolute value of  spectrum \eqref{spectrum},  we note that  in the presence of the phase disorder,  the spectrum amplitude exhibits a small-scale noise.
 Although the overall shape of the spectrum amplitude  is reproducible, its fine details may vary between experimental realizations due to the noise.
 This noise can be suppressed by averaging the spectrum amplitude  over experimental realizations.
 In analytical derivation we consider that the averaging is performed over the infinite number of realizations. 
In experiments,  
as numerical simulations demonstrate, 
40--50 realizations usually suffice to strongly suppress the noise.
Moreover, for sufficiently long condensate chains, one may perform spatial averaging by dividing the system into segments of equal length, evaluating the spectrum amplitude independently within each segment, and subsequently averaging the results.
 To simplify the derivation of the analytical formula,   we  square  the spectrum amplitude before the averaging.
Taking into account that for $\sigma \ll d$ 
the different terms  in  sum \eqref{spectrum}  overlap  weakly in the  $k$-space, 
  we obtain
\begin{multline}
\langle
|\tilde{n}(k,t)|^2
\rangle
 = N_{0}^2 e^{-k^2\sigma^2}  \\ 
 \times \sum\limits_{n = 1-M}^{M-1} \exp\left(  -\frac{d^2}{4\sigma^2}   \left( \frac{ktd}{\pi T_d}  - n \right)^2    \right)\langle \left| f_n(k) \right|^2 \rangle .
  \label{spectrum_abs}
\end{multline}

 As given by equation \eqref{spectrum_abs},    the dependence of the average square of the spectrum amplitude $ \langle |\tilde{n}(k,t)|^2 \rangle $ on the initial phases  is determined  by the function $\langle \left| f_n(k) \right|^2 \rangle$.  
This function    can be  expressed in terms of the phase correlation function $\langle e^{i (\varphi_j - \varphi_p)} \rangle = \alpha( | j - p | )$:
\begin{widetext}
\begin{equation} 
 \langle \left|  f_n(k) \right|^2 \rangle  
=\alpha^2(|n|)\frac{\sin ^2 (kd(M-|n|)/2)}{\sin ^2 (kd/2)} + \\ \sum\limits_{p=1-|n|}^{|n|-1} (M-|n| - |p|)\cos(pkd) [\alpha^{2}(|p|) - \alpha^{2}(|n|)]  .
\label{f}
\end{equation}
In  \eqref{f}, it is implied that  $\sum\limits_{i=a}^b = 0$ if $a>b$.
\end{widetext}

\begin{figure*}
\center
\includegraphics[scale=0.8]{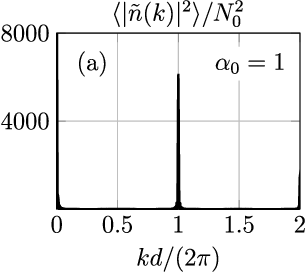}
\hspace{0.1cm}
\includegraphics[scale=0.8]{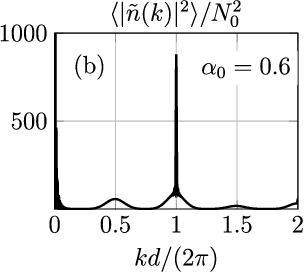}
\hspace{0.1cm}
\includegraphics[scale=0.8]{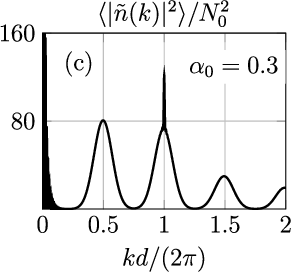}
\hspace{0.1cm}
\includegraphics[scale=0.8]{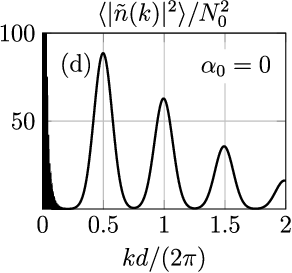}
\caption{\small 
Average square of spectrum amplitude $\langle |\tilde{n}(k,t)|^2 \rangle$ for a  chain of  $M=100$   Bose--Einstein condensates at time $t = T_d$,  plotted using equations \eqref{spectrum_abs} and \eqref{f}.
The  initial phases of the condensates are (a) identical,   (b),  (c) partially disordered,  (d)  completely disordered.
}
\label{fig:analytical_spectrum}
\end{figure*}

When the initial phases are identical,  
$\alpha (|p|)=1$ $\forall p \in \{0,  \pm 1, \dots, \pm (M-1) \}$,  and  the function $\langle \left| f_n(k) \right|^2 \rangle$ takes the following form:
\begin{equation} 
 \langle \left|  f_n(k) \right|^2 \rangle  
= \frac{\sin ^2 (kd(M-|n|)/2)}{\sin ^2 (kd/2)} .
\label{f1}
\end{equation}
Therefore,  the spectrum amplitude consists of the  peaks at positions  $ k =  \frac{2\pi n}{d} $ $\left(  n = 0, \pm 1, \dots, \pm (M-1) \right)$.
In contrast,   when the phases are completely disordered,   $\alpha (|p|)=0$ $\forall p \in \{\pm 1, \dots, \pm (M-1) \}$,  and 
\begin{equation}
 \langle \left|  f_n(k) \right|^2 \rangle  
=
 \begin{cases}
   M- |n| &\text{ for $n\neq 0$, } \\
    \frac{\sin ^2 (kdM/2)}{\sin ^2 (kd/2)} &\text{for $ n =0$.}
 \end{cases}
 \label{f2}
\end{equation}
In this case,  according to equations  \eqref{spectrum_abs} and \eqref{f2},  
 the peaks $ k =  \frac{2\pi n}{d} $ with  $   n = \pm 1, ..., \pm (M-1) $  disappear.
Instead, new peaks emerge in the spectrum amplitude at wave numbers $ k = \frac{\pi n}{d} \frac{T_d}{t}$ $\left( n = \pm 1, ..., \pm (M-1) \right)$,  which depend  on time.
Figures \ref{fig:analytical_spectrum}(a) and (d) show the mean squares of the spectrum amplitude computed by formulas \eqref{spectrum_abs}, \eqref{f1},  and \eqref{f2} for identical and fully disordered initial phases, respectively.
Due to the symmetry 
 $\langle |\tilde{n}(k,t)|^2 \rangle  = \langle |\tilde{n}(-k,t)|^2 \rangle $,    we present the spectrum only for  non-negative values of $k$.
As the figures demonstrate, the disorder-induced spectrum peaks are significantly broader than those   associated with identical phases.

When the initial phases are partially disordered, 
the exact shape of the average square of the spectrum amplitude $\langle |\tilde{n}(k,t)|^2 \rangle$ is determined by the specific form of the phase correlation function $\alpha( |p|)$  $\left( p \in \mathbb{Z}  \right)$.  
If the phase disorder is produced by thermal fluctuations,
 the phase correlation function   decays exponentially with the distance:  $\alpha ( |p|) = \alpha_0^{|p|}$ \cite{BoseChainFluctPitaevskii2001}.
Here,   $ \alpha_0  = \langle e^{i(\varphi_j - \varphi_{j+1})} \rangle$ is the coherence factor;  it determines the  degree of the initial phase disorder.
Examples of  the  average square of the spectrum amplitude $\langle |\tilde{n}(k,t)|^2 \rangle$   for the thermal phase fluctuations with  $\alpha_0 = 0.6$ and $\alpha_0 = 0.3$ at time $t=T_d$  are shown  in  Figs. \ref{fig:analytical_spectrum}(b) and (c).
According to the figures, when the initial phases are partially disordered,
the spectrum amplitude  contains both the narrow peaks at $ k =  \frac{2\pi n}{d} $ $\left(  n = \pm 1, ..., \pm (M-1)   \right)$ (arising from the residual phase coherence) and the broad peaks at  $ k = \frac{\pi n}{d} \frac{T_d}{t}$   $\left(  n = \pm 1, ..., \pm (M-1)    \right)$ (associated with the phase disorder). 
The ratio of the heights of these two types of peaks depends on the value of the coherence factor $\alpha_0$.

It is worth to notice that,  for  completely disordered initial phases, in the  long chain limit $M\rightarrow \infty$,  an analytical expression for  the spectrum  was obtained in Ref.~\cite{RandomPhaseInteference2019}.
According  to \cite{RandomPhaseInteference2019},  
\begin{equation}
\tilde{n}(k,t) \propto \frac{2\pi}{d}\delta(k) + \frac{\sqrt{\pi M}}{2} e^{-k^2\sigma^2} \sum\limits_{j=-\infty, j\neq 0}^{\infty}  e^{-\left( j - \frac{ktd}{T_d\pi}\right)^2} e^{i\varphi_j'},
\label{spectrumPRL}
\end{equation}
where $\varphi_j'$ are random phases satisfying condition $\varphi_j' = -\varphi_{-j}'$.
In contrast to expressions  \eqref{spectrum_abs},  \eqref{f2},   formula \eqref{spectrumPRL} presents the spectrum itself without taking the squared absolute value of it.
Formula \eqref{spectrumPRL} can be derived from our  expressions \eqref{spectrum} and \eqref{f0} by taking the long-chain limit and neglecting the small-scale noise.

\section{Physical interpretation of the spectrum peaks  \label{sec:explanation} }

We now clarify the origin of the spectrum peaks that emerge when the initial phases  become disordered.
To this end,   we consider the formula for the one-dimensional density that arises from taking   the squared absolute value of  wave function \eqref{wave_function}
 \begin{widetext}
\begin{equation}
  n_{1\text{D}}(z,t) =   \sum\limits_{j=1}^M n_j(z, t)
   + 2 \sum\limits_{1 \leq p < j \leq M} \underbrace{ \sqrt{n_j(z, t) n_p(z,t)} \cos \left(   k_{j-p}(t)z + \Phi_{j,p}(t) \right)}_{ \substack{ n_{j,p}(z,t)  \text{ --- wavelet resulting from}   \text{   interference of the } j\text{-th and }p\text{-th condensates}} } ,
   \label{3cond}
   \end{equation}    
    \end{widetext}
    where
     \begin{equation}
      k_{j-p}(t) = \frac{\pi T_d}{td } (j-p) , 
 \label{1d_wave_number}
\end{equation}
     \begin{equation}
 \Phi_{j,p}(t)=- k_{j-p}(t) \frac{(j+p)d}{2}  - (\varphi_j - \varphi_p) ,
 \label{1d_phase}
\end{equation}
 $n_j(z,t) = |\psi_j(z,t)|^2$ is the density of the $j$-th condensate with $\psi_j(z,t)$  being the wave function  of the $j$-th condensate.
The expression for the wave numbers  $k_{j-p}(t)$ is derived under the assumption $\omega t  \gg 1 $, 
which  is valid at times when the condensates overlap and interfere.

Density of the condensate chain \eqref{3cond} is composed of harmonic waves which are localized within a finite spatial region.
These waves  are commonly referred to as wavelets~\cite{WaveletsChui1992}.
Each wavelet $n_{j,p}(z,t )$ originates from the interference of   the  $j$-th and $p$-th condensates, as follows from  the density formula for a pair of condensates
\begin{eqnarray}
\left| \psi_{j}(z,t) + \psi_{p}(z,t)\right|^2 = n_{j}(z,t) + n_{p}(z,t) 
\nonumber \\  +  2  \underbrace{ \sqrt{n_{j}(z,t)n_{p}(z,t)}\cos \left(   k_{j-p}(t)z + \Phi_{j,p}(t) \right)}_{n_{j,p}(z,t )} .
\end{eqnarray}
The wavenumber $k_{j-p}(t)$ of the wavelet $n_{j,p}(z,t )$ is determined by the separation between the condensates  $j-p$  from whose interference it originates.

It is worth noting that  the wave number $k_n(t)$ of the density wavelets  coincides with the position of the $n$-th disorder-induced  peak in  spectrum amplitude  \eqref{spectrum_abs}.  
This observation naturally suggests that the  $n$-th spectral peak originates from the wavelets with the wave numbers  $k_n(t) $.
To  test this hypothesis,  we extract  from  density formula \eqref{3cond} the sum of the wavelets with such wave numbers
\begin{equation}
\mathcal{N}_n (z,t )   =     \sum_{p=1}^{M-n}  n_{p+n,  p}(z,t), 
\end{equation}
 and  examine how the phase disorder affects it.
For  simplicity,  we neglect the spatial localization of the individual wavelets,  as its inclusion yields  qualitatively the same result.
Under this approximation,   the sum $\mathcal{N}_n(z, t)  $  reduces,  up to a prefactor,  to a sum of harmonic waves:
 \begin{multline}
 \mathcal{N}_n (z,t )   \propto   \sum_{l=1}^{M-n}  \cos \left[ k_{n}(t) z - k_{n}(t) d l - k_{n}(t) n/2  \right. 
  \\ 
 - \left. \varphi_{l+n} + \varphi_l \right] .
 \label{harmonic_waves}
 \end{multline}

Harmonic waves with identical wave numbers  
combine  into a harmonic wave with the same wave number.
Therefore,    formula  \eqref{harmonic_waves}  can be  represented as
\begin{equation}
\mathcal{N}_n (z,t )   \propto  A_n \cos( k_n z + B_n ), 
\end{equation}
 where $A_n = \left| C_n \right|  $ with  $  C_n =  \sum_{l=1}^{M-n} e^{i(k_n l d  + \varphi_{n+l} - \varphi_l)} $ is  the resultant amplitude,    $B_n = \arctan ( \text{Im} (C_n) / \text{Re} (C_n))$  is the overall phase shift.
The amplitude of the resultant wave depends on the phases of the summed waves.
In the absence of the phase disorder,  when  $k_n(t)d = 2\pi p$ for some $  p  \in \mathbb{N} $,    all the individual waves in  sum \eqref{harmonic_waves}  are in phase.
Therefore,   the waves  amplify each other during the summation,
and the resultant amplitude   reaches its maximum $A_n = M-n \gg 1$.
When $  k_n(t)d  $ deviates from $2\pi p$, the phases of the waves differ. 
Specifically, each subsequent wave in  sum \eqref{harmonic_waves} differs from the previous one in phase by $ -  k_n(t)d  $.
As a consequence of these phase differences, for any $  l  $-th wave in  sum \eqref{harmonic_waves},  there exists an $  (l+q)  $-th wave such that its phase differs from that of the $  l  $-th wave by approximately $  \pi  $, where $  q  $ satisfies $  q k_n(t) d \approx  \pi + 2\pi s  $ for some $s \in\mathbb{N}$. 
Thus, for nearly every wave in  sum \eqref{harmonic_waves}, there is another wave with an approximately opposite phase.
Waves with opposite phases cancel each other during summation. 
Therefore, the amplitude $  A_n  $ becomes significantly suppressed $  A_n \ll M-n  $.
This mutual cancellation of waves is confirmed by the exact expression for the amplitude
\begin{equation}
 A_n(t)  =  \left| \frac{\sin \left(k_l(t)d(M-l)/2 \right)}{\sin \left(k_l(t)d/2 \right)} \right|,
 \label{A_n}
\end{equation}
according to which $A_n = M-n$ when $k_l(t)d = 2\pi  $,  and $A_n \ll M-n$ when $k_l(t)d  $ deviates from $2\pi p  $.
The amplification and cancellation of the waves during their summation are commonly referred to as constructive and destructive interference, respectively.
In general,  the harmonic wave $A_n\cos( k_n z + B_n )$ contributes  to the two peaks in the amplitude spectrum with positions $ k = \pm k_n (t)$ and heights proportional to  $A_n$.
The presence of the sharp maxima of the amplitude  $A_n$  when $k_n(t)d = 2\pi p$ causes the spectrum peaks $ k = \pm k_n (t)$ to effectively reduce to the peaks  $ k = \pm 2\pi p /d$.

 The constructive and destructive interference of the waves in  sum \eqref{harmonic_waves} vanish when the initial phases of the condensates become disordered.
This occurs because the disorder in the initial phases randomizes the phases of the harmonic waves in  sum \eqref{harmonic_waves}.
Therefore, under the condition $  k_n(t)d = 2\pi p  $ $\left( p \in \mathbb{N}\right) $, the phase alignment of the harmonic waves is lost.
Likewise, when $k_n(t)d \neq 2\pi p $, the waves are no longer in antiphase,
and their mutual cancellation vanishes.
Due to the absence of both the constructive and destructive interference,  the amplitude $  A_n  $ of the resultant wave becomes independent of the specific value of $  k_n(t)  $.
Calculation of the average amplitude  $ \langle A_n \rangle $ under completely disordered initial phases gives the following value
\begin{equation}
 \langle A_n \rangle =  \sqrt{ \pi( M-n) / 4} 
\end{equation}
   for  arbitrary  $k_n(t)$.
Therefore,  the sum of the wavelets $ \mathcal{N}_n (z,t )$ gives rise to the  peaks at $k = \pm k_n(t)$ in the spectrum amplitude.
Since each wavelet in the sum  $ \mathcal{N}_n (z,t )$  results from the interferences of the condensate pairs initially separated by the distance  $nd$,
it follows that the spectral peaks $k = \pm k_n(t)$ arise specifically from these pairwise interferences.

\section{Extension to an arbitrary condensate lattice \label{sec:general_lattices}  }

In addition to the  condensate chains, the Talbot effect can also be observed in two- and three-dimensional condensate lattices of various geometries.
Trivial examples of 2D and 3D lattices  exhibiting the Talbot effect are the square and cubic lattices, which are straightforward higher-dimensional generalizations of the chain.
Other lattice geometries  exhibiting the Talbot effect can be adopted from optics.
For example,   in optics,  the Talbot effect has  been demonstrated for the   hexagonal \cite{Rogers1962, Winthrop1965} and circular \cite{Badalyan2013}  arrays of wave sources.

We now show that the peaks  corresponding to the pairwise interferences of the condensates appear in the spatial density spectrum  in response  to the phase disorder for any condensate lattice, regardless of its geometry.
To do this, we consider the expression for the spatial density distribution of an arbitrary lattice of freely expanding condensates 
\begin{widetext}
\begin{equation}
  n(\mathbf{r},t) =   \sum\limits_{j=1}^M n_j(\mathbf{r}, t) 
   +  \sum_{\substack{j,p=1\\
                  j\neq p}}^M  \sqrt{n_j(\mathbf{r}, t) n_p(\mathbf{r},t)} \cos \left(   \mathbf{k}_{j,p}(t) \mathbf{r} + \Phi_{j,p}(t) \right),
   \label{gen_density}
   \end{equation}    
  \end{widetext}
   where
     \begin{equation}
\mathbf{k}_{j,p}(t) = \frac{m}{\hbar t} \left( \mathbf{R}_j-\mathbf{R}_p \right) , 
 \label{3d_wave_number}
\end{equation}
     \begin{equation}
 \Phi_{j,p}(t)=- \frac{1}{2} \mathbf{k}_{j,p}(t) \left( \mathbf{R}_j+\mathbf{R}_p \right)  - (\varphi_j - \varphi_p) .
 \label{3d_phase}
\end{equation}
Equations \eqref{gen_density},  \eqref{3d_wave_number} and \eqref{3d_phase} follow from taking the squared absolute value of the wave function of a freely expanding condensate lattice.
The initial positions of the condensates  are specified by the vectors $\mathbf{R}_j$.
The condensates are assumed to have identical widths in all spatial directions.

In expression \eqref{gen_density}, the spatial density distribution is written as a sum of the density wavelets generated by the pairwise interferences of the condensates.
When the condensate phases are disordered,  
the phases of the wavelets are disordered as well, 
 as follows from  \eqref{3d_phase}.
As  demonstrated in the previous section,  
  wavelets with  identical wave vectors and disordered phases,  when superposed,  
produce a peak in the amplitude spectrum at that wave vector.
Since expression \eqref{gen_density}  contains  the wavelets with  many different wave vectors $\mathbf{k}_{j,p}(t)$,  the phase disorder produces multiple spectral peaks, each arising from a set of the wavelets sharing the same wave vector.
In contrast, when the initial phases are identical, the peaks with wave vectors $\mathbf{k}_{j,p}(t)$ are absent due to the constructive and destructive interference among the overlapping wavelets.

\begin{figure*}
\begin{minipage}{1.0\linewidth}
\includegraphics[height=1.8in]{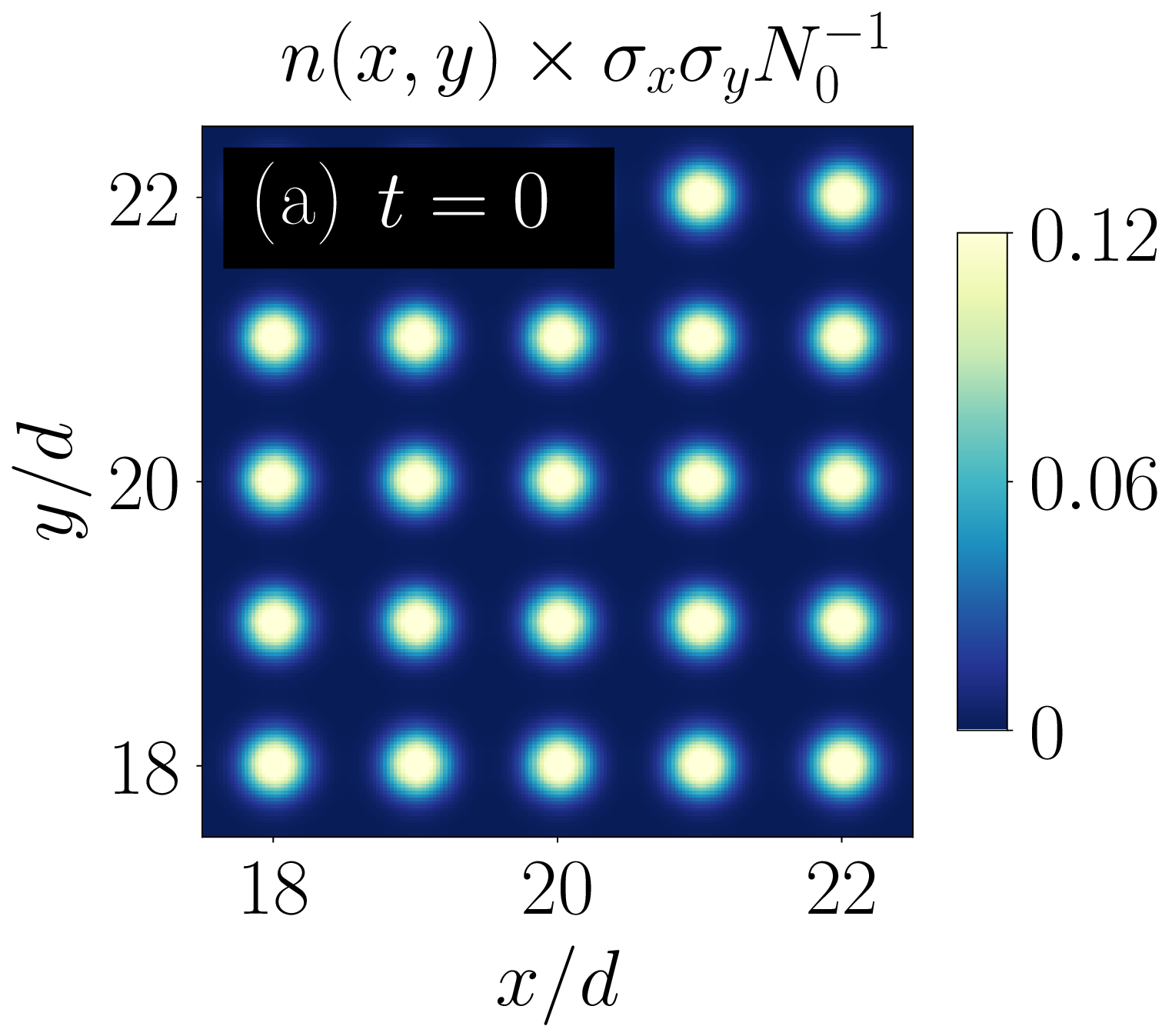}
\hspace{0.4cm}
\includegraphics[height=1.8in]{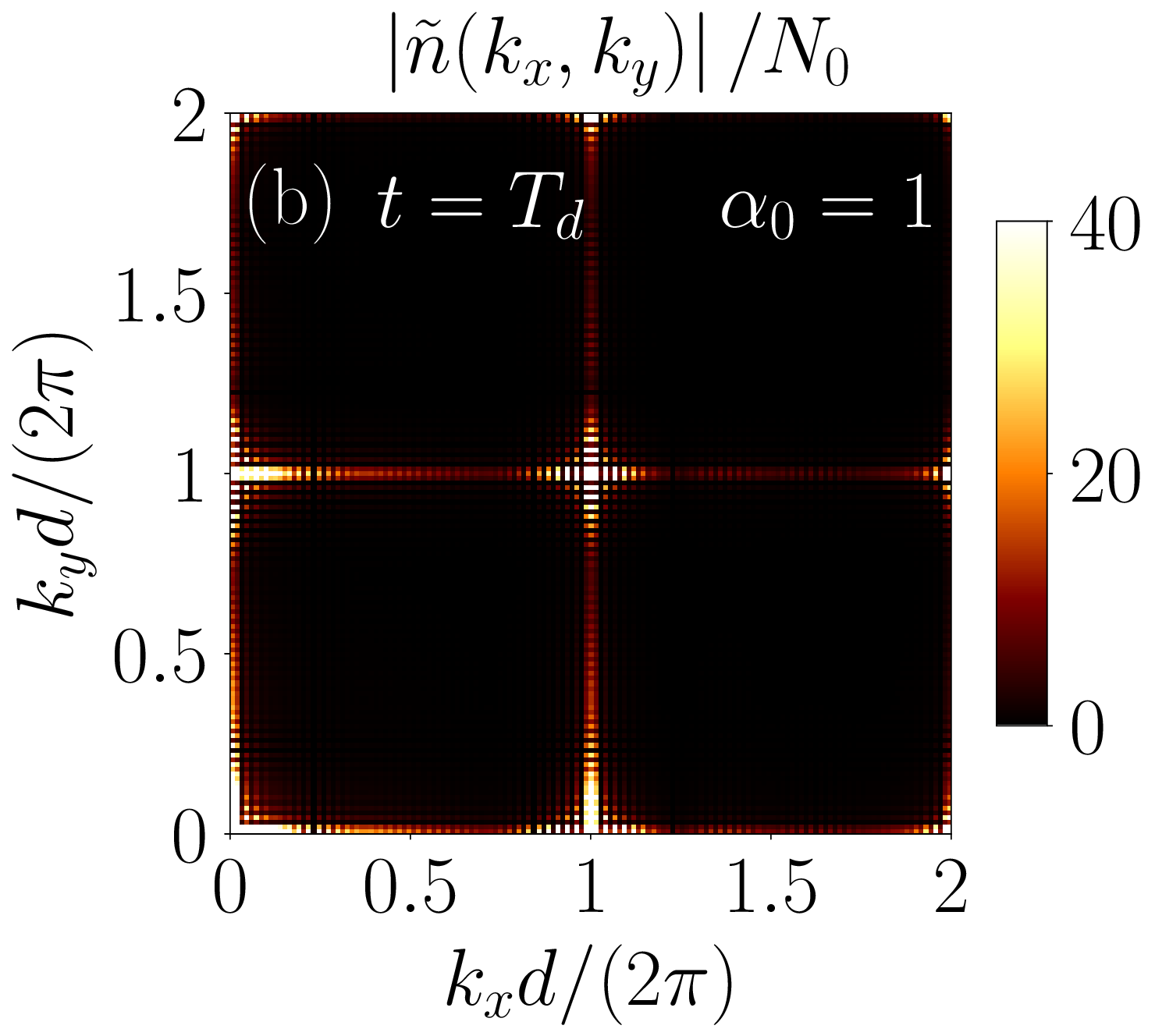}
\hspace{0.4cm}
\includegraphics[height=1.8in]{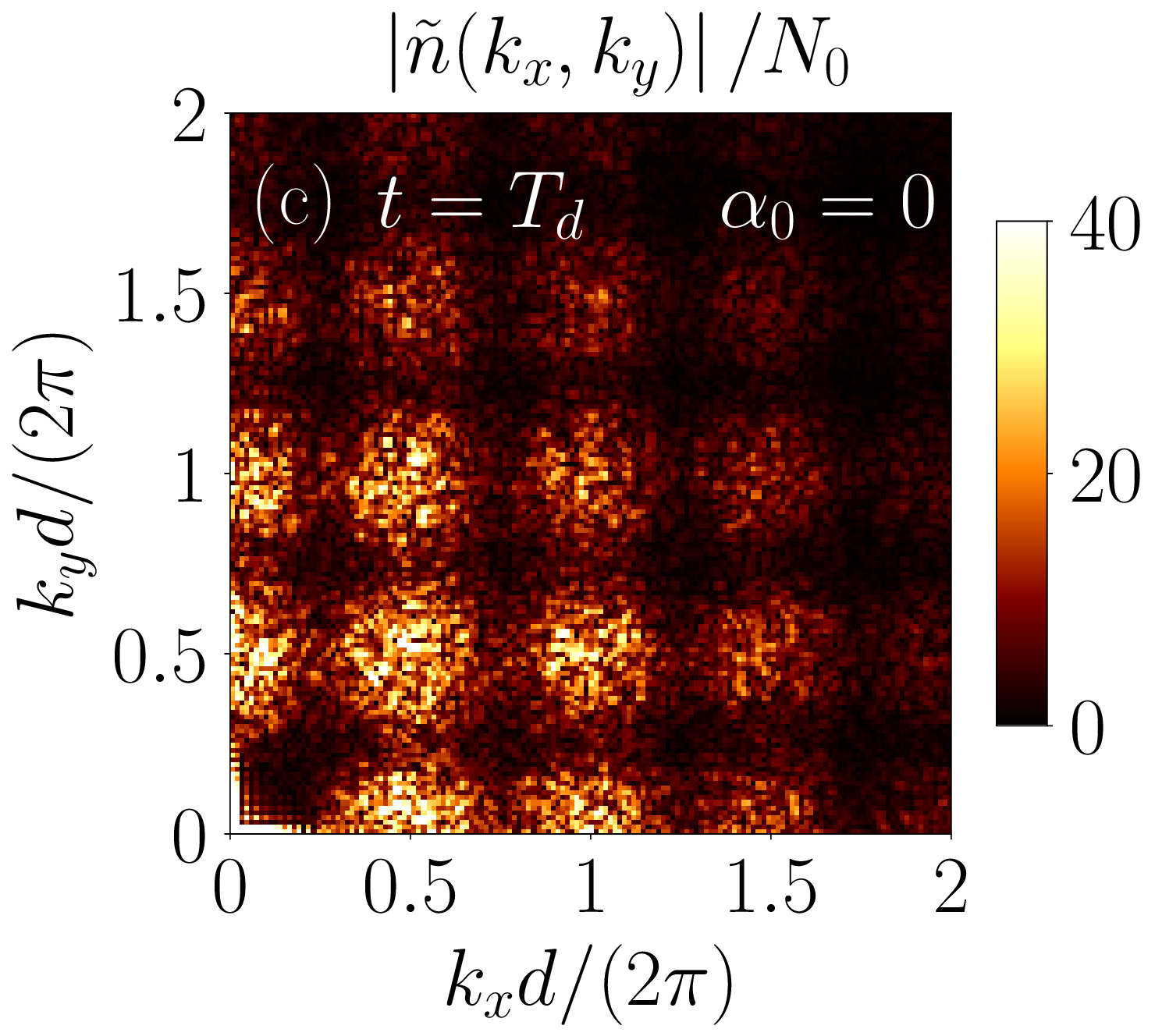}
\end{minipage}

\vspace{0.3cm}
\begin{minipage}{1.0\linewidth}
\center
\includegraphics[height=1.8in]{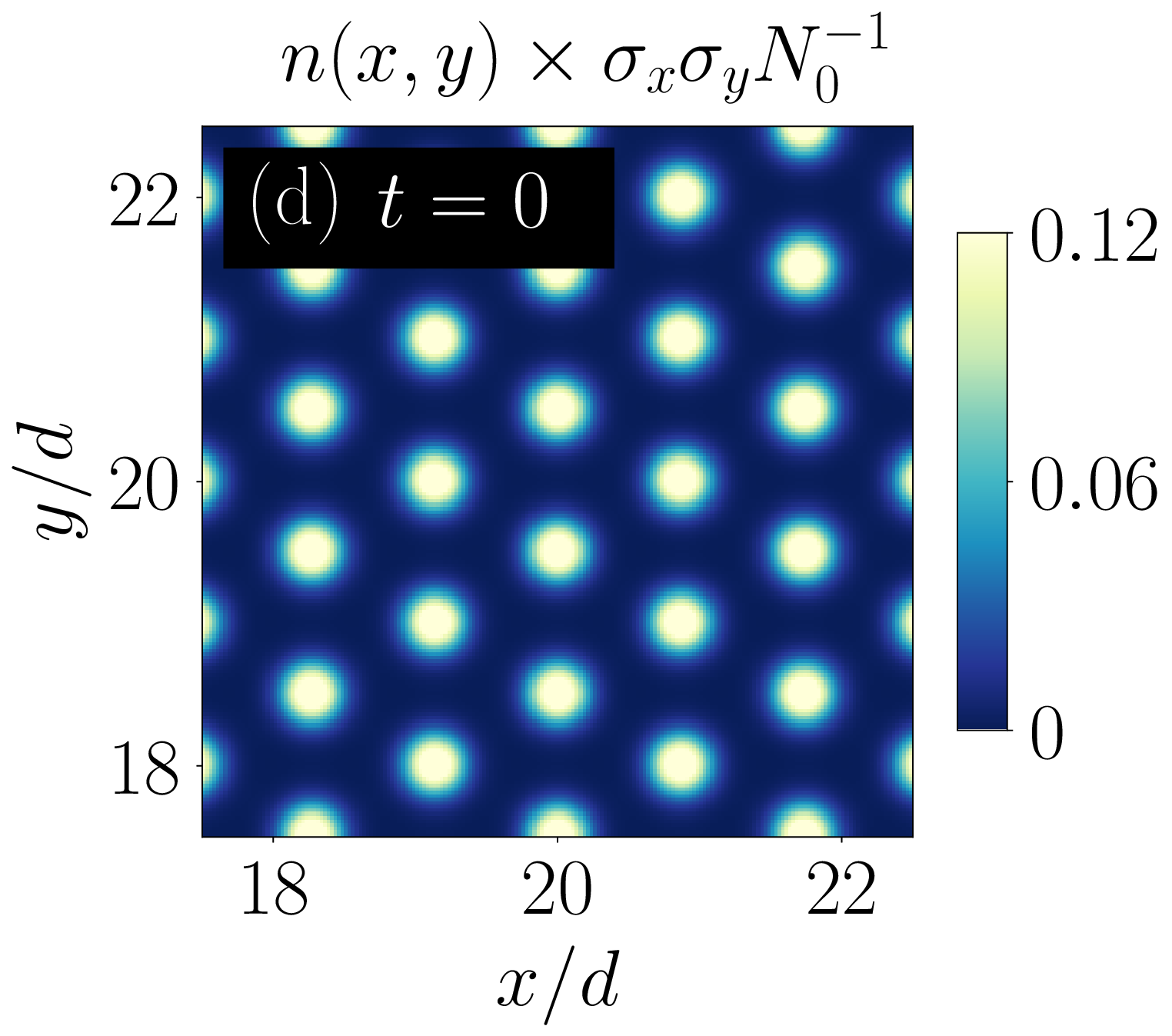}
\hspace{0.4cm}
\includegraphics[height=1.8in]{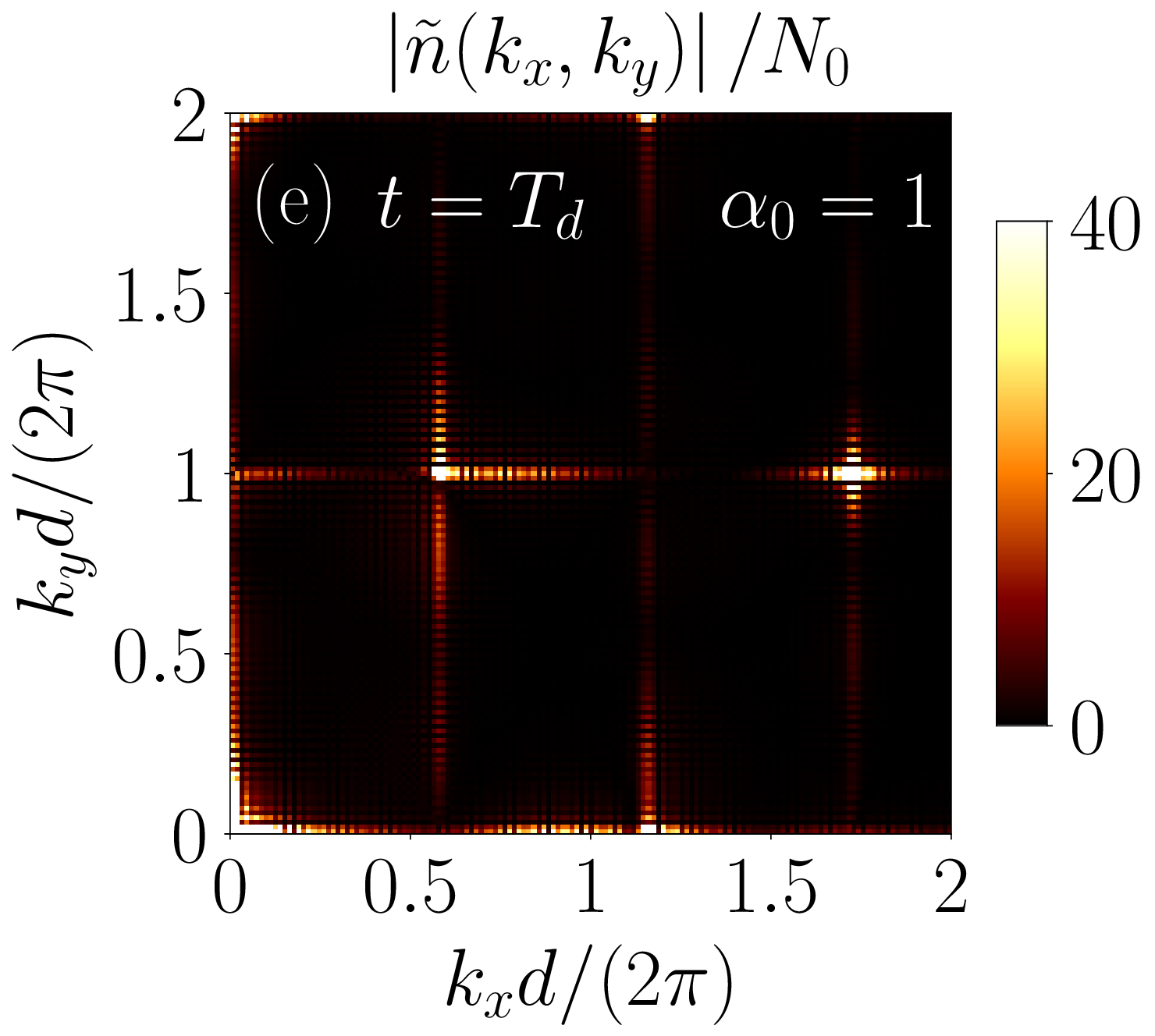}
\hspace{0.4cm}
\includegraphics[height=1.8in]{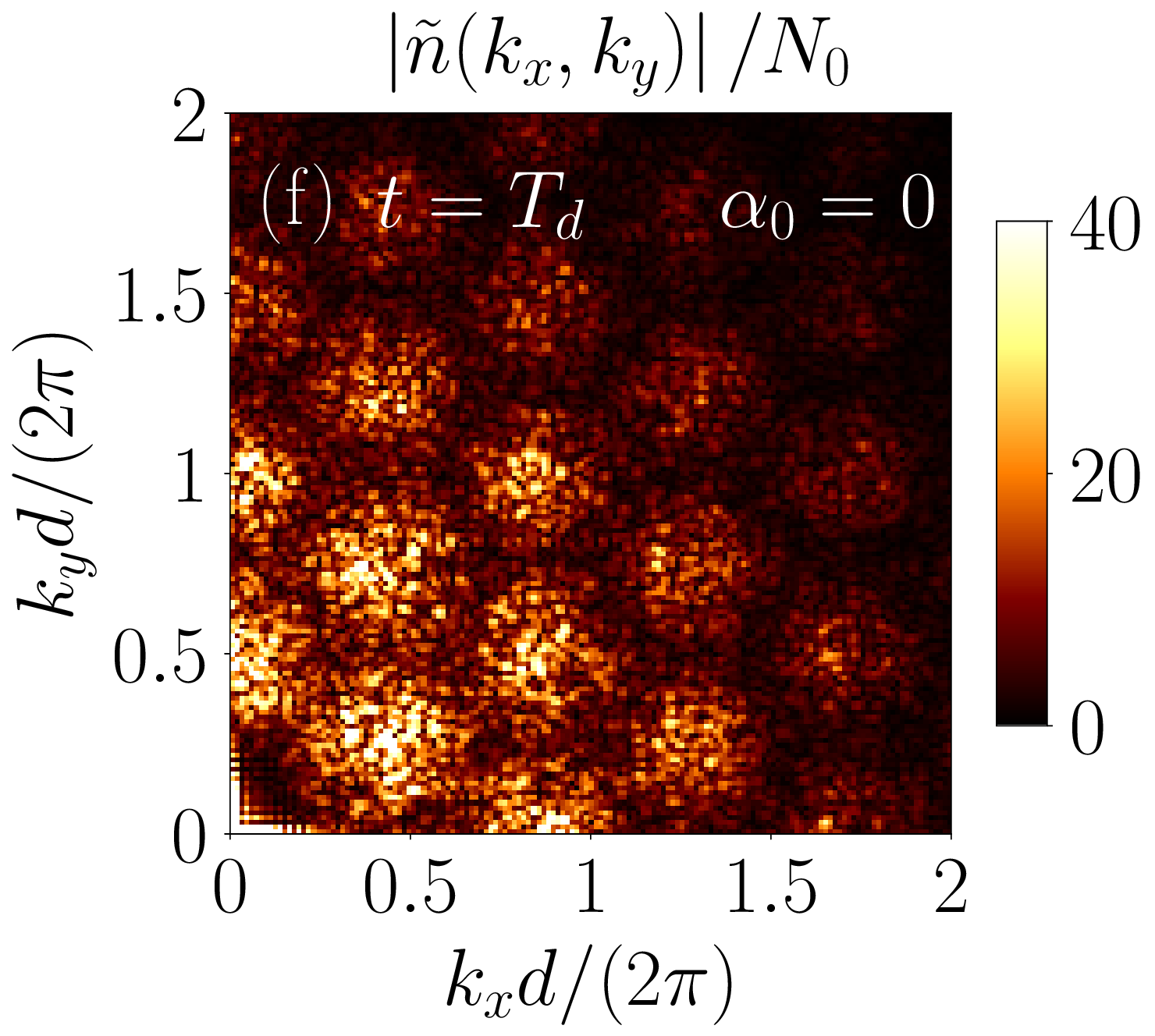}
\end{minipage}
\caption{
Interference of the square and hexagonal lattices of Bose--Einstein condensates.
The initial lattice configurations are depicted in figures  (a),  (d);  total sizes of the lattices are  40 $\times$ 40.
For  identical initial condensate phases,  the  amplitudes  of the spatial density spectra at time $t=T_d$  are shown in figures  (b),  (e).
For  completely disordered phases,  the amplitudes  of the spatial density spectra at $t=T_d$ are displayed in figures  (c),  (f).   }
\label{fig:2d_lattice}
\end{figure*}

The  geometry of the condensate lattice determines the arrangement of the peaks generated by the phase disorder in the  spectrum amplitude.
According to   \eqref{3d_wave_number},  the wave vector  $\mathbf{k}_{j,p}(t)$ of the density wavelet resulting from   the interference between the $j$-th and $p$-the condensates is defined by their relative position $\mathbf{R}_j - \mathbf{R}_p$.
Therefore,    the  spectral peaks  induced  by the  phase disorder  replicate the relative  positions of the condensates  in the initial lattice configuration.
As an illustration,  we plot in Fig.   \ref{fig:2d_lattice}   the amplitude of the spatial density spectrum  $\left| \tilde{n}(k_x,k_y) \right|$ for the square and hexagonal condensate lattices.
According to the figure,  
for the square lattice with  completely disordered initial phases, the spectrum amplitude  exhibits a square lattice of interference peaks.
Similarly, the hexagonal lattice yields a  hexagonal arrangement of  peaks in the spectrum amplitude.

\section{Comparison with the Fraunhofer diffraction  \label{sec:Fraunhofer}}

The qualitative change of the spectrum amplitude in response to the  initial phase disorder occurs only in the Fresnel diffraction regime
where the Talbot effect  takes place. 
In the Fraunhofer diffraction regime,  
this qualitative change disappears, and  the effect  of the phase disorder on the spectrum amplitude  becomes  quantitative. 
To demonstrate the vanishing of the qualitative change of the spectrum amplitude,  we consider Eqs.~\eqref{spectrum_abs} and \eqref{f} in the Fraunhofer regime.
Since in the Fraunhofer regime
 the positions of the Gaussian peaks in \eqref{spectrum_abs} $k = n \frac{\pi}{d}\frac{T_d}{t}$ are much smaller than $2\pi / (d (M-|n|))$,  the function $\frac{\sin ^2(kd(M-|n|)/2)}{\sin^2(kd/2)}$ in \eqref{f} can be replaced by $(M-|n|)^2$,  and $\cos(pkd)$ can be replaced by $1$.
Thus, expression \eqref{f} becomes
\begin{multline}
 \langle \left|  f_n(k) \right|^2 \rangle  
=\alpha^2(|n|)(M-|n|)^2  \\
+  \sum\limits_{p=1-|n|}^{|n|-1} (M-|n| - |p|) [\alpha^{2}(|p|) - \alpha^{2}(|n|)]  .
\label{f_far_field}
\end{multline}

For identical initial phases, $\alpha (|p|)=1$ $\forall p \in \{0,  \pm 1, \dots, \pm (M-1) \}$,  and expression \eqref{f_far_field} simplifies to
\begin{equation}
 \langle \left|  f_n(k) \right|^2 \rangle  
=(M-|n|)^2   .
\label{f_far_field1}
\end{equation} 
Combined with Eq.~\eqref{spectrum_abs}, this implies that,  in the Fraunhofer regime with identical initial phases,  the spectrum consists of distinct peaks at $  k = n \frac{\pi}{d} \frac{T_d}{t}  $ ($  n = 0, \pm 1, \dots, \pm (M-1)  $), each with amplitude scaling as $  M^2  $. 
This is in striking contrast to the Fresnel regime where identical phases produce no such  peaks at these wave vectors.

In the opposite limit of completely disordered initial phases, $  \alpha(|p|) = 0  $ for all $  p \neq 0  $, and the function $ \langle \left|  f_n(k) \right|^2 \rangle  $ becomes
\begin{equation}
 \langle \left|  f_n(k) \right|^2 \rangle  
=
 \begin{cases}
   M- |n| &\text{ for $n\neq 0$, } \\
    (M-|n|)^2 &\text{for $ n =0$}.
 \end{cases}
 \label{f_far_field2}
 \end{equation}
 Thus,  spectrum peaks still appear at the same positions $  k = n \frac{\pi}{d} \frac{T_d}{t}  $.
 The only difference with the identical phases case lies in the heights of the spectrum peaks. 
 Under completely disordered phases,  only the central ($  n=0  $) peak has height scaling as $  M^2  $, while all side peaks ($  |n| \geq 1  $) scale linearly as $  M  $.

\begin{figure*}
\center
\includegraphics[scale=0.8]{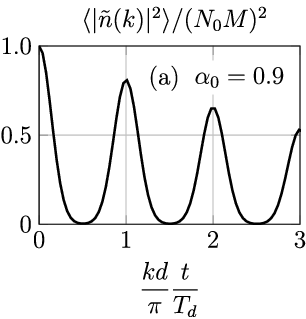}
\hspace{0.5cm}
\includegraphics[scale=0.8]{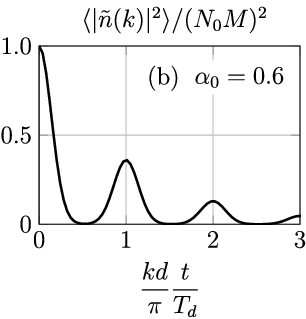}
\hspace{0.5cm}
\includegraphics[scale=0.8]{ 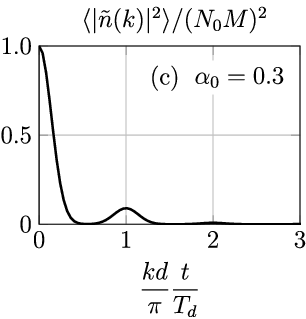}
\caption{\small 
Average square of spectrum amplitude $\langle |\tilde{n}(k,t)|^2 \rangle$ for a  chain of  $M=100$   Bose--Einstein condensates at time $t = 500T_d$ in the Fraunhofer regime. 
}
\label{fig:spectrum_far_field}
\end{figure*}

When the initial phases are partially disordered, the relative heights of the spectrum peaks are  governed by the phase correlation function $  \alpha(|n|)  $. 
A simple expression for the peak heights can be obtained in the limit of a long condensate chain. 
In this limit,  in Eq.   \eqref{f_far_field},  the terms linear in $  M  $ are negligible compared to the $  M^2  $ contribution, yielding $  \langle \left| f_n(k) \right|^2 \rangle = \alpha^2(|n|) M^2  $. 
Therefore, the height of the $  n  $-th peak equals $N_0^2 e^{-k_n^2\sigma^2} M^2 \alpha^2(|n|)$ with $k_n = \frac{\pi n}{d} \frac{T_d}{t}$.
This expression for the peak heights was first derived and experimentally tested in Ref.~\cite{Wang2012}. 
As an illustration,  under the long chain approximation,  we  plot the average squared spectrum amplitude $  \langle |\tilde{n}(k,t)|^2 \rangle  $ in Fig.~\ref{fig:spectrum_far_field}.
In the plot,  we consider the phase fluctuations to be thermal;  thus,   $  \alpha(|n|) = \alpha_0^{|n|}  $, where  $  0 \leq \alpha_0 \leq 1  $ \cite{BoseChainFluctPitaevskii2001}.

Another  difference between the Fresnel and Fraunhofer diffraction regimes concerns the contrast of the interference fringes.
In the Fraunhofer regime, the fringe contrast is finite for a condensate chain of finite length and vanishes in the limit of an infinitely long chain. 
The disappearance of interference fringes in the long-chain limit was derived in Ref. \cite{Hadzibabic2004} and follows from the expression for the spatial density distribution in the Fraunhofer regime
\begin{multline}
n_{1\text{D}}(z,t) \propto M e^{-z^2/(2\sigma^2\omega^2 t^2)} \\ 
\times \left(1 + \sum\limits_{n=1}^{M-1} D_n \cos\left( E_n + n \frac{\pi }{d} \frac{T_d}{t} z  \right) \right),
\label{far_field}
\end{multline}
where $D_n$,  $E_n$ is the absolute value and the argument of $2/ M \sum_{j=N+1}^{M}e^{i ( \varphi_j - \varphi_{j-n})}  $.
The fringe contrast of the $n$-th harmonic is directly proportional to $D_n$.
For completely disordered initial phases,  Ref.  \cite{Hadzibabic2004} showed that
$ \langle D_n^2 \rangle = 4(M-n)/M^2 $, which tends to zero in the long-chain limit ($M \to \infty$).

In the Fresnel regime, the fringe contrast is independent of the chain length $M$. 
This property originates from the locality of condensate wave-packet overlap. 
In the Fresnel regime,  only those condensate wave packets whose initial separation does not exceed  $2\sigma \omega t $ overlap during the expansion. 
As a result, the density modulation at a given position $z$ is determined solely by condensates whose expanded wave packets overlap at that point, i.e., by condensates that were initially located within a distance of order $\sigma\omega t$ from $z$.
 Condensates outside this region contribute negligibly to the local interference signal.
 Consequently, increasing the total chain length $M$ does not modify the fringe pattern itself; it only extends the spatial region over which the fringes are observed.

  \section{Summary   \label{sec:conclusion}  }

We have examined the influence of  initial phase disorder  on the Talbot effect in  a lattice of freely expanding Bose--Einstein condensates.
We have demonstrated that the phase disorder induces a qualitative change in the spectrum of the spatial density distribution of the condensates.
Specifically, the spectrum acquires  peaks that are absent when the initial phases are identical.
We have shown that these peaks result from the pairwise interferences of the condensates.
The positions of these peaks coincide with the wave vectors of the density wavelets generated by the pairwise interferences.
The absence of such peaks  under identical initial phases is explained  by  the mutual destruction of the overlapping wavelets.

For a one-dimensional condensate chain,  we have derived  an analytical expression for  the  spectrum of the spatial density distribution   for  an arbitrary phase disorder.
According to this formula, under partially disordered initial phases, the spectrum contains both the peaks arising from the pairwise interferences of the condensates and those corresponding to the uniform initial phases.
The relative heights of these two types of peaks   are governed by the degree of the phase disorder.
Under completely disordered initial phases,  the peaks  corresponding to uniform initial phases disappear.


\nocite{*}

\bibliography{bib}

\end{document}